\documentclass[aps,superscriptaddress,floatfix]{revtex4}

\usepackage{amssymb}
\usepackage{latexsym}
\usepackage{graphics}
\usepackage{graphicx}
\usepackage{dcolumn}
\usepackage{bm}
\usepackage{color}
\newcommand{\gammap}{\dot\gamma}

\begin{document}

\title{Recent experimental probes of shear banding}
\author{S{\'e}bastien Manneville}
\email{sebastien.manneville@ens-lyon.fr}
\affiliation{Laboratoire de Physique, Universit\'e de Lyon -- \'Ecole Normale Sup\'erieure de Lyon  -- CNRS UMR 5672\\46 all\'ee d'Italie, 69364 Lyon cedex 07, France}

\begin{abstract}
Recent experimental techniques used to investigate shear banding are reviewed. After recalling the rheological signature of shear-banded flows, we summarize the various tools for measuring locally the microstructure and the velocity field under shear. Local velocity measurements using dynamic light scattering and ultrasound are emphasized. A few results are extracted from current works to illustrate open questions and directions for future research.
\end{abstract}

\maketitle

\section{Introduction}
\label{intro}
Under simple shear conditions some complex fluids may separate into bands of widely different viscosities. This phenomenon is known as ``shear banding'' and involves inhomogeneous flows where macroscopic bands bearing different shear rates or shear stresses coexist in the sample. Although shear banding is usually attributed to an underlying shear-induced transition from a given microscopic organization of the fluid structure to another, it still raises lots of theoretical and experimental challenges.

Shear banding has been suspected or evidenced in a growing number of complex materials ranging from aqueous surfactant solutions, foams, granular slurries to polymers, liquid crystals, or concentrated colloidal suspensions and emulsions. Among all these materials ``wormlike micelle'' solutions have emerged as a model system to study shear banding. Depending on the concentration, most of these self-assembled surfactant systems constituted of long, cylindrical, semi-flexible aggregates undergo a shear-induced transition from a viscoelastic state of entangled, weakly oriented micelles to a state of highly aligned micelles above some critical shear rate $\dot{\gamma}_1$. Such a transition is strongly shear-thinning since the viscosity of the aligned state can be orders of magnitude smaller than the zero-shear viscosity of the system. Under simple shear and above $\dot{\gamma}_1$, the system spatially separates into coexisting bands of high and low viscosities $\eta_1$ and $\eta_2$ corresponding respectively to the entangled and aligned states. As the shear rate is increased above $\dot{\gamma}_1$, the shear-induced structure progressively expands along the velocity gradient direction at constant shear stress $\sigma=\sigma_c$, until the system is fully aligned at some shear rate $\dot{\gamma}_2$. Thus the rheological signature of shear banding is the existence of a horizontal plateau at $\sigma_c$ in the shear stress vs shear rate constitutive curve $\sigma(\dot{\gamma})$, which extends from $\dot{\gamma}_1$ to $\dot{\gamma}_2$.

\begin{figure}[htbp]
\scalebox{0.8}{\includegraphics{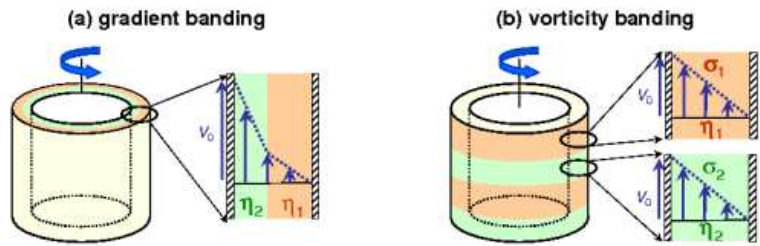}}
\caption{Shear banding (a) in the gradient direction and (b) in the vorticity direction in Couette geometry.}
\label{f.rheol}
\end{figure}

The simple shear banding scenario described above is referred to as ``gradient banding'' in the literature. Another situation known as ``vorticity banding'' may also occur, where the system separates at a constant shear rate into bands bearing different stresses stacked along the vorticity direction, corresponding to a vertical portion in the flow curve, {\it i.e.}, to a shear-thickening transition. Figure~\ref{f.rheol} illustrates the gradient and vorticity banding situations as expected in Couette geometry. Moreover in some gels and in more concentrated materials known as ``soft glassy materials'' such as emulsions, pastes, or colloidal suspensions, the transition may be from a solid state to a liquid state in connection with the yield stress phenomenon. In such cases the flow may be constituted of a flowing band that coexists with a solid region. Although such a solid--liquid transition (or ``unjamming'' transition) can be seen as the limit case of the previous gradient banding scenario where $\gammap_1\rightarrow 0$, it raises a number of experimental issues concerning thixotropy, ageing, wall slip, fracture, and the measurement of very small shear rates and velocities. Here we shall thus only briefly mention recent results on soft glassy systems. Finally let us mention that the term shear banding is also used in the field of plasticity in amorphous solids. For the sake of conciseness this report will be mainly restricted to gradient banding in self-assembled surfactant systems. 

The aim of the present paper is to review the various experimental tools that have emerged in the last decade to assess shear-banded flows. In section~\ref{rheo} we first recall the rheological signature of shear banding, namely the existence of a plateau in the flow curve, and give examples of various systems where such a signature was found. We show that deeper characterizations of both the local structure and the velocity field under shear are needed to fully investigate shear banding. Section~\ref{structure} deals with the local characterization of the fluid microstructure. Section~\ref{velo} then lists the various velocimetry techniques that may be used to characterize the velocity field of shear-banded flows, with emphasis on two techniques developed at Centre de Recherche Paul Pascal in Bordeaux, namely dynamic light scattering and ultrasonic velocimetry. Finally in section~\ref{questions} we extract a few results from the recent literature that highlight some open questions and give directions for future research.

\section{The rheological signature of shear banding and the need for local rheophysical approaches}
\label{rheo}

As explained in the introduction the rheological signature of gradient (resp. vorticity) shear banding is the presence of a horizontal stress plateau (resp. vertical portion) in the flow curve $\sigma(\dot{\gamma})$. The first experimental evidence for a stress plateau in nonlinear rheological measurements on wormlike micellar systems was provided by Rehage and Hoffmann \cite{Rehage:1991} on the CPyCl--NaSal system. Further research effort established the generality of this peculiar feature on other wormlike micelle systems \cite{Berret:1997a,Berret:1998b,Berret:1994a,Cappelaere:1997,Hassan:1996,Mendez:2003a,Soltero:1999,Yesilata:2006}. A recent review of both linear and nonlinear rheological properties of wormlike micelles is available in ref.~\cite{Berret:2005} which also addresses shear banding in concentrated and semidilute micellar systems.

Stress plateaus (or quasi-plateaus) have been observed in the nonlinear rheological response of a wide variety of other self-assembled systems such as lyotropic lamellar phases and multilamellar vesicles \cite{Bonn:1998,Roux:1993,Wunenburger:2001}, thermotropic liquid crystals \cite{Panizza:1995}, lyotropic hexagonal phases \cite{Ramos:2000}, liquid crystalline polymers \cite{Pujolle:2001,Pujolle:2002}, diblock copolymers \cite{Holmqvist:2002}, triblock copolymers \cite{Waton:2004}, soft cubic crystals \cite{Eiser:2000}, or telechelic polymers \cite{Berret:2001a}. Reference~\cite{Hamley:2000} provides a review on the effect of shear in block copolymer solutions. Semidilute and entangled solutions of high molecular weight polymers  \cite{Bercea:1993,Hu:2007,Inn:2005,Islam:2001,Mhetar:2000,Pattamaprom:2001,Tapadia:2003,Tapadia:2004} and polymer melts \cite{Boukany:2006,Drda:1995,Piau:1994,Piau:1995} may also present an almost flat flow curve within a given range of shear rates. 

Evidence for shear-thickening behaviours that could be linked to shear-induced transitions and vorticity shear banding was first reported in dilute wormlike micelle solutions in ref.~\cite{Rehage:1982} and later in refs. \cite{Berret:2000,Berret:1998a,Boltenhagen:1997a,Boltenhagen:1997b,Fischer:2002,Hartmann:1997,Hu:1998b,Hu:1998a,Hu:1994,Hu:1993,Liu:1996,Oda:1997}. Shear-thickening is also found in lyotropic lamellar phases during the formation of multilamellar vesicles known as ``onions'' in the literature \cite{Bergenholtz:1996,Diat:1993b,Roux:1993,Wilkins:2006}. In all these cases strong shear-thickening is observed over a narrow range of shear rates that separates a quasi-Newtonian behaviour from a shear-thinning branch in the flow curve.

Thus the two types of shear-banded flows are expected from nonlinear rheological measurements. However it remains to be checked whether or not the flow is actually inhomogeneous, {\it i.e.}, whether or not spatial separation into bands bearing different viscosities occurs as sketched in Fig.~\ref{f.rheol}. Indeed rheological measurements only provide data ($\gammap$, $\sigma$, $\eta$) which are averaged over the whole sample. These {\it global} data are also referred to as {\it engineering} data in the literature. Looking for a plateau in the flow curve is clearly insufficient to evidence shear banding and {\it local} measurements of both the organization of the microstructure and the flow field are required to ensure that the flow becomes inhomogeneous. In particular, in the framework of the simple gradient banding scenario described in the introduction, the validity of the so-called ``lever rule'' which gives the proportion $\alpha$ of the shear-induced state as a function of the shear rate $\dot{\gamma}$ along the stress plateau~\cite{Spenley:1993}:
\begin{equation}
\label{eq_lr} \dot\gamma = (1-\alpha) \dot\gamma _1 +\alpha \dot\gamma _2,
\end{equation}
should be investigated experimentally.\footnote{It is not our purpose to discuss theories of shear banding in the present experimental review. Theoretical details will be found in the corresponding reviews in the same volume.} Therefore developing {\it spatially resolved} tools to study shear-banded flows has become the subject of intense effort in the last decade.

Moreover it has been shown experimentally that shear-induced transitions are usually associated with much more complex phenomena than simple horizontal or vertical portions in the flow curve. These phenomena include:

\noindent(i) slow rheological transients at the onset of the stress plateau in shear-thinning systems \cite{Berret:1997b,Berret:1999,Berret:1994a,Berret:1994b,Cappelaere:1995,Decruppe:2001,Grand:1997,Mendes:1997,Porte:1997},

\noindent(ii) huge induction times (typically minutes to hours) for shear-thickening behaviours \cite{Berret:2000,Boltenhagen:1997a,Boltenhagen:1997b,Hu:1998a,Liu:1996} and/or formation of the shear-induced structure \cite{Courbin:2002,Courbin:2003,Escalante:2000,Leon:2000,Panizza:1998,Zipfel:1999},

\noindent(iii) hysteretic flow curves \cite{Bonn:1998,Mendez:2003a,Panizza:1995,Ramos:2000,Volkova:1999,Wunenburger:2001},

\noindent(iv) oscillating rheological responses \cite{Fischer:2000a,Grosso:2003,Herle:2005,Hilliou:2002,Wheeler:1998,Wunenburger:2001},

\noindent(v) chaotic-like rheological responses, known as ``rheochaos'' \cite{Bandyopadhyay:2000,Bandyopadhyay:2001,Ganapathy:2006,Pimenta:2006,Pujolle:2003,Salmon:2002}, and

\noindent(vi) successive transitions that do not necessarily involve well-defined plateaus \cite{Azzouzi:2005,Eiser:2000,Roux:1993,vanderGucht:2006}.

These complex {\it temporal} behaviours point to metastable states, bistability, phase transitions, and bifurcations between various microstructural organizations and/or flow regimes. Although very informative, purely rheological studies do not allow to discriminate between shear banding and other possible space- and time-dependent phenomena that may occur in complex fluids such as wall slip \cite{Barnes:1995,Denn:2001}, stick-slip \cite{Boukany:2006,Drda:1995,Pujolle:2003}, fracture \cite{Berret:2001b,Hu:1998b,Inn:2005,Liu:1996}, tumbling in nematic materials \cite{Berret:1995a}, or elastic instabilities \cite{Groisman:1998,Groisman:2000,Larson:1990,McKinley:1996,Muller:1989,Shaqfeh:1996}. Consequently, investigating shear banding requires more thourough characterizations using {\it spatially and temporally resolved} tools. In the following we first present structural probes that have been used recently to evidence shear banding. We then focus on velocimetry techniques under shear. These various experimental tools are usually combined with rheometry and are referred to as ``rheophysical'' techniques.

\section{Structural probes of shear-banded flows}
\label{structure}

Structural probes aim at answering the following questions:

\noindent(i) Can the coexistence of various microstructural organizations be evidenced and followed under shear in order to confirm or invalidate the simple scenario for gradient banding described in the introduction and summarized by eq.~(\ref{eq_lr})?

\noindent(ii) What is the nature of the shear-induced structure?

\noindent(iii) Can a local order parameter be defined?

\noindent(iv) Can concentration fluctuations be evidenced and can the coupling between flow and concentration be studied?

\subsection{Rheo-optical tools}

Rheo-optics, namely birefringence and turbidity measurements under shear, are a convenient way to evidence the inhomogeneous structure of a shear-banded flow when the optical properties of the microstructure at low shear and the shear-induced microstructure differ significantly.

\subsubsection{Birefringence studies}

The first evidence for flow inhomogeneities and coexistence along the stress plateau in shear-thinning wormlike micellar systems came from flow birefringence studies in Couette geometry by Decruppe {\it et al.} \cite{Berret:1997a,Cappelaere:1997,Cappelaere:1995,Decruppe:1995,Decruppe:2003,Makhloufi:1995}. It was shown that a highly birefringent band nucleates at the onset of the stress plateau and grows as the shear rate is increased between $\gammap_1$ and $\gammap_2$. The birefringence intensity and the extinction angle were measured in the birefringent band as a function of the imposed shear rate. Assuming that the birefringent band corresponds to the low viscosity material, the ``lever rule'' (eq.~(\ref{eq_lr})) has been successfully tested on various systems \cite{Azzouzi:2005}.

However time-resolved birefringence measurements have also revealed a number of unexpected features such as the existence of sub-bands, the absence of connection between the stress relaxation and the growth of the birefringent band, and the presence of three bands of different optical properties in Couette geometry \cite{Lerouge:2000,Lerouge:1998,Lerouge:2004,Miller:2007}. Three-band structures and temporal fluctuations were also reported in ref.~\cite{Lee:2005}. Such features have questioned the simple picture of two shear bands separated by a sharp interface. 

Birefringence is of course not limited to wormlike micelle solutions and has also been used to investigate, {\it e.g.}, flow-induced isotropic--nematic transitions in liquide-crystalline polymers \cite{Mather:1997} and vorticity banding in rodlike viruses \cite{Kang:2006,Lettinga:2004,Lettinga:2005}.

\subsubsection{Turbidity and scattered light measurements}

Turbidity measurements and direct imaging of scattered light have been mostly used to investigate the shear-thickening transition in dilute or equimolar wormlike micelles \cite{Fischer:2002,Wheeler:1998}. Although these techniques are more qualitative, they allow to follow the build-up of the shear-induced structures \cite{Boltenhagen:1997a,Boltenhagen:1997b,Hu:1998b,Hu:1998a,Liu:1996}. 
Surprisingly, some turbidity and scattered light results coupled to rheometry and birefringence measurements suggest that the flow during shear-thickening could be inhomogeneous not only in the vorticity direction but also in the gradient direction \cite{Berret:2002,Herle:2005}.

Finally recent scattered light imaging experiments in Couette geometry have shown the possibility of an undulating interface between shear bands in a gradient banding wormlike micellar system \cite{Lerouge:2006}.

\subsection{Scattering techniques}

Scattering techniques under shear provide a very useful tool to get more quantitative data on the structure of the shear-induced phase. Depending on the characteristic length scale of the microstructure and on its sensitivity to radiation, one may use small angle light scattering (SALS), small angle neutron scattering (SANS), or small angle X-ray scattering (SAXS). The main drawback of all these techniques is that they do not usually provide local information but rather some scattering pattern that is integrated over the whole volume sampled by the incident beam. 

\subsubsection{Light scattering}
SALS has been used to probe a wide variety of self-assembled systems on typical length scales of 0.1--100~$\mu$m. In particular, SALS measurements yield the size and the arrangement of micrometric multilamellar vesicles during the shear-induced lamellae--onions transition and during the layering transition from disordered to ordered onions in some lyotropic lamellar phases \cite{Bergenholtz:1996,Courbin:2002,Courbin:2004,Diat:1995a,Muller:1999,Panizza:1998,Sierro:1997}.

Highly anisotropic SALS patterns were reported in wormlike micelles and interpreted in terms of an isotropic--nematic transition and a flow-induced string phase \cite{Kadoma:1996,Kadoma:1998}. SALS may be also convenient to investigate shear-enhanced concentration fluctuations in wormlike micellar systems. Such concentration fluctuations typically show up as butterfly patterns or bright streaks in the SALS patterns \cite{Herle:2005,Kadoma:1997,Schubert:2004,Waton:2004}. 

\subsubsection{Neutron scattering}
In wormlike micelle solutions, SANS has proven very efficient to evidence shear-induced isotropic--nematic transitions in concentrated systems  \cite{Berret:1994b,Cappelaere:1997,Schmitt:1994}, shear-induced alignment in more dilute systems \cite{Jindal:1990}, shear-induced phase separation similar to that observed in polymers \cite{Schubert:2004}, and shear-induced vesicle--wormlike micelle transitions \cite{Mendes:1997}. Crescent-like SANS patterns are the signature of elongated micelles aligned in the velocity direction. Such patterns allow one to define an order parameter
(although these estimates are spatially averaged) and to measure the proportion of each coexisting phase as well as their concentration \cite{Berret:1998b,Roux:1995}. In particular, the results of ref.~\cite{Berret:1998b} show significant deviations from the simple gradient banding scenario such as the presence of a sloped stress ``plateau,'' the failure of the lever rule, and possible variations of the viscosities $\eta_1$ and $\eta_2$ with the global shear rate. These deviations were attributed to flow--concentration coupling.

Similar SANS measurements under shear have been performed on lyotropic lamellar phases \cite{Diat:1993b,Escalante:2000,Zipfel:1999}, diblock copolymers \cite{Hamley:1998,Koppi:1993}, and liquid-crystalline polymers where a shear-induced smectic-{\it{A}}--smectic-{\it{C}} transition was found \cite{Noirez:2000,Noirez:1997}. The reader is also referred to ref.~\cite{Hamley:2000} for a review of SANS and SAXS measurements on block copolymer solutions under shear. However, so far, such measurements have been unable to spatially resolve the two coexisting phases due to large beam widths or to strong absorption. Very recently, local SANS experiments have been performed for the first time that allow to study the order parameter in a Couette geometry of gap 1.35~mm with a spatial resolution of 200~$\mu$m \cite{Liberatore:2006}. The results in ref.~\cite{Liberatore:2006} were obtained on a wormlike micellar system that shows shear-induced phase separation. They suggest that the shear-induced phase is composed of a highly branched, concentrated micellar solution coexisting with a nearly isotropic, brine phase.

Finally SANS was used to characterize the shear-induced structure in shear-thickening transitions. Scattering patterns were interpreted as the superposition of a low-viscosity state made of short agregates and an entangled sheared network of longer micelles \cite{Berret:1998a}. Time-resolved SANS also allowed one to probe the long-time relaxation of the shear-thickened gel-like state, showing that the shear-induced structure can remain stable for several hours \cite{Oda:2000}.

\subsubsection{X-ray scattering}
SAXS measurements under shear have been performed to evidence shear-induced orientational ordering in various self-assembled systems such as thermotropic liquid crystals \cite{Panizza:1995}, lyotropic hexagonal phases \cite{Ramos:2000}, diblock copolymers \cite{Hamley:2000,Pople:1998a,Pople:1997}, and soft cubic phases \cite{Eiser:2000,Molino:1998}. Thanks to narrow beam widths of typical diameter 0.2~mm, X-rays allow for spatially resolved measurements across the gap of a Couette cell. For instance ref.~\cite{Eiser:2000} shows unambiguously the spatial coexistence of different orientation states under shear in correlation with a stress plateau in the flow curve of a soft cubic phase. Note also that spatially-resolved X-ray diffraction has also been recently used to measure the intermembrane spacing in a sheared lamellar system and to infer velocity profiles in a pipe flow \cite{Welch:2002}.

\subsection{Nuclear Magnetic Resonance}

Nuclear Magnetic Resonance (NMR) can be performed under shear to assess locally the microstructure of a complex material through the measurement of the order parameter using deuterium NMR spectroscopy, of molecular diffusion from pulsed gradient spin echo NMR, and of local concentrations. A specific review focusing on rheo-NMR applications to shear banding is available in the present volume. The reader is thus referred to ref.~\cite{Callaghan:2008} for full technical details and references on microstructural characterization using NMR under shear.

\section{Characterizing the velocity field of shear-banded flows}
\label{velo}

Structural characterization under shear has confirmed the occurence of shear-induced transitions associated with the peculiar rheological behaviours described in section~\ref{rheo}. Local measurements have also evidenced the spatial coexistence of different organizations of the microstructure. However pointwise velocity measurements are still needed to check whether the shear rate profile is indeed constituted of bands of different viscosities. This section is thus devoted to a short review of velocimetry techniques that have been used to characterize the flow field during rheological experiments in potentially shear-banding materials. Such techniques should provide answers to the questions below:

\noindent (i) Does the shear rate profile present shear bands?

\noindent (ii) Are the velocity profiles consistent with the lever rule (eq.~(\ref{eq_lr}))?

\noindent (iii) Do these bands coincide with birefringence bands?

\noindent (iv) Is the flow field stationary? If not, does an ``instantaneous'' version of eq.~(\ref{eq_lr}) hold?

\subsection{NMR velocimetry}

NMR can be used not only for structural characterization but also for velocity measurements. Using NMR velocimetry, Callaghan {\it et al.} have provided the first clear evidence for inhomogeneous shear rate profiles in wormlike micelles in pipe flow \cite{Callaghan:1996}, Couette geometry \cite{Mair:1996,Mair:1997}, and cone-and-plate geometry \cite{Britton:1997,Britton:1999}. Later works combining NMR velocimetry and deuterium NMR spectroscopy have shown that birefringence bands do not necessarily correspond to shear bands \cite{Fischer:2000b,Fischer:2001}. Finally, more recently, fast NMR velocimetry allowed the authors to resolve temporal fluctuations of the flow field \cite{Holmes:2003a,Lopez:2006,Lopez:2004}

Since the early work by Callaghan {\it et al.}, NMR velocimetry has been used in a growing number of complex fluids to investigate shear banding or shear localization including emulsions \cite{Gotz:2003}, colloidal systems \cite{Holmes:2004,Wassenius:2005}, wet granular materials \cite{Huang:2005}, cement pastes \cite{Jarny:2005}, and other soft glassy materials studied by Coussot {\it et al.} \cite{Bertola:2003,Coussot:2002,Ragouilliaux:2006,Raynaud:2002}.

The typical spatial resolution of current NMR velocimetry is about 50~$\mu$m over gaps of about 1~mm. Full velocity profiles can be recorded roughly every second although information on much smaller time scales of the order of milliseconds can also be gained from pointwise velocity distributions \cite{Holmes:2003a}. The main drawback of NMR velocimetry is that simultaneous measurements of global stress and local velocity are generally impossible due to the presence of strong magnets. Thus rheological measurements have to be performed using a different apparatus and correlating temporal fluctuations of the flow field to those of the rheological variables is not possible. Again the reader will find more details about NMR results on shear banding in the companion review article \cite{Callaghan:2008}.

\subsection{Particle tracking velocimetry} 

Although generally restricted to transparent materials seeded with small tracer particles, particle tracking velocimetry (PTV) and particle image velocimetry (PIV) have been applied to the study of shear banding flows in combination with rheometry. PIV was first used in Couette geometry on dilute solutions of cationic surfactants \cite{Koch:1998}: a one-dimensional but strongly fluctuating velocity profile was evidenced in the shear-thickening region of the flow curve with apparent wall slip and a local maximum, suggesting that the shear-induced ``gel'' is macroscopically heterogeneous.

The capillary flow of semidilute CPyCl--NaSal wormlike micelles was later studied through PIV by M{\'e}ndez-S{\'a}nchez {\it et al.} \cite{Mendez:2003b}. It was shown that apparent slip at the walls occurs even before the occurence of ``spurt'', {\it i.e.}, before the viscosity drop by an order of magnitude along the stress plateau. Plug-like and shear-banded velocity profiles were evidenced and allowed the authors to reconstruct the ``true'' flow curve from local measurements.

In a remarkably thorough study, Hu and Lips used time-resolved PTV in Couette geometry with a 2~mm gap to follow the kinetics of shear banding in the CPyCl--NaSal system \cite{Hu:2005}. Steady-state velocity profiles showed no wall slip and revealed a shear banding scenario that seem roughly consistent with the lever rule (although the local shear rates $\gammap_1$ and $\gammap_2$ seem to vary a bit with the imposed shear rate and although the fully aligned state could not be accessed in these experiments due to flow instabilities). The spatial resolution is about 10~$\mu$m and the temporal resolution is 0.1~s (time interval between two successive velocity profiles). An even faster temporal resolution (2.5~ms per profile) with a 100~$\mu$m spatial resolution over a 6~mm gap was achieved by Miller and Rothstein who also focused on the semidilute CPyCl--NaSal system  and confronted transient velocity profiles to flow-induced birefringence measurements \cite{Miller:2007}.

Very recently time-resolved PTV was applied to entangled polymer solutions under shear \cite{Hu:2007,Tapadia:2006b,Tapadia:2006a,Wang:2006a} to address the issues of the monotonicity of their constitutive curve, the possibility of shear banding at least during transients, and the presence of edge fracture \cite{Inn:2005,Tapadia:2003,Tapadia:2004}. The flow of many other complex materials has also been investigated by PIV or PTV including two-dimensional foams \cite{Debregeas:2001,Janiaud:2005,Kabla:2003,Lauridsen:2004,Lauridsen:2002,Wang:2006b}, two-dimensional dry granular materials \cite{Mueth:2003,Mueth:2000}, soft particle pastes \cite{Meeker:2004a,Meeker:2004b}, rodlike viruses \cite{Kang:2006}, and granular pastes \cite{Barentin:2004,Lenoble:2005}. Finally it should be noted that particle tracking can be achieved in concentrated colloidal suspensions by using fast confocal microscopy so that jamming, yielding, banding, and slip phenomena are now accessible with microscopic spatial resolution \cite{Besseling:2007,Isa:2007,Isa:2006}.

\subsection{Photon correlation spectroscopy}

The shear rate profile or the velocity profile in weakly scattering samples can be determined from photon correlation spectroscopy techniques such as dynamic light scattering (DLS) in homodyne or heterodyne modes. 

\subsubsection{Homodyne DLS} If one considers the auto-correlation function $C(\tau)=<I(t)I(t+\tau)>$ of the light intensity $I(t)$ scattered from a given scattering volume inside a sheared material, one can show that $C(\tau)$ falls off with a characteristic time that is inversely proportional to the shear rate $\gammap$ in the scattering volume \cite{Fuller:1980,Salmon:2003b}. More precisely, knowing the exact shape of the scattering volume and the intensity profile of the incident beam, one may perform ``pointwise'' measurements of the local shear rate, as shown in ref.~\cite{Wang:1994} in the case of a polymer solution. However such a determination of the local velocity gradient using homodyne DLS is subject to a number of restrictions due to the nature of the scattering objects and to the flow geometry \cite{Salmon:2003b}, so that heterodyne setups may be preferred for combined rheological and flow measurements.

\subsubsection{Heterodyne DLS} In heterodyne DLS, the scattered light is mixed to a reference beam so that $C(\tau)$ reflects the interferences between the Doppler shifted light that has crossed the sheared sample and the reference light. In that case, it can be shown that $C(\tau)$ oscillates with the frequency $\mathbf{q}\cdot\mathbf{v}$, where $\mathbf{q}$ is the scattering wave vector. Thus recording the autocorrelation function \cite{Salmon:2003b} or the Fourier transform of the intensity resulting from the interferences \cite{DiLeonardo:2005} leads to an estimate of the average velocity inside the scattering volume, once $\mathbf{q}$ is known either from the geometry of the experiment or from a calibration procedure.

\begin{figure}[htbp]
\scalebox{0.8}{\includegraphics{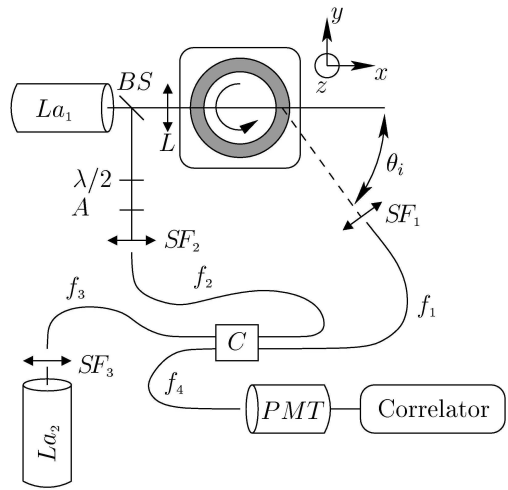}}
\caption{Heterodyne DLS setup. $La$ denotes lasers, $BS$ a beam splitter, $\lambda/2$ a half-wave plate, $A$ a neutral density filter, $SF$ spatial filters, $C$ the device coupling optical fibers $f$, and $PMT$ the photomultiplier tube. Reprinted from ref.~\cite{Salmon:2003b}.}
\label{f.dls}
\end{figure}

\begin{figure}[htbp]
\scalebox{0.8}{\includegraphics{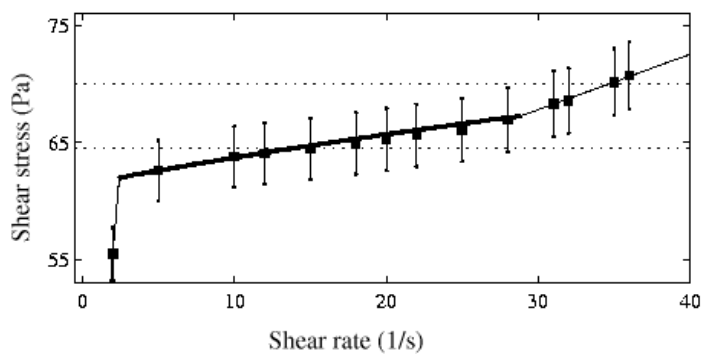}}
\caption{Steady-state engineering flow curve for a 6\%~wt. CPyCl--NaSal solution in 0.5~M brine recorded in a 1~mm gap Couette cell under imposed shear rate. The thick continuous line between the two branches of the flow curve is the prediction of the simple gradient banding scenario with $\gammap_1=2.5$~s$^{-1}$, $\gammap_2=26$~s$^{-1}$, and $\sigma_c = 64.2$~Pa. The slope of the stress plateau is attributed to the curvature of the Couette cell. The horizontal lines correspond to the local values of the shear stress at the rotor at the beginning and at the end of the coexistence regime. The vertical bars indicate the variation of the local shear stress across the gap. Reprinted from ref.~\cite{Manneville:2004c}.}
\label{f.cpcl1}   
\end{figure}

Figure~\ref{f.dls} shows a sketch of the optical fiber based heterodyne DLS setup developed by the Bordeaux group around a commercial rheometer equipped with a Couette geometry \cite{Salmon:2003b}. The spatial resolution is fixed by the size of the scattering volume and is about 50~$\mu$m. Depending on the scattered intensity, the autocorrelation function is accumulated over typically 1~s to yield one ``pointlike'' velocity measurement. To construct velocity profiles, the scattering volume is simply scanned across the gap by mechanically moving the rheometer, which takes about 1~min per profile. Full technical details can be found in ref.~\cite{Salmon:2003b}.

\begin{figure}[htbp]
\scalebox{0.8}{\includegraphics{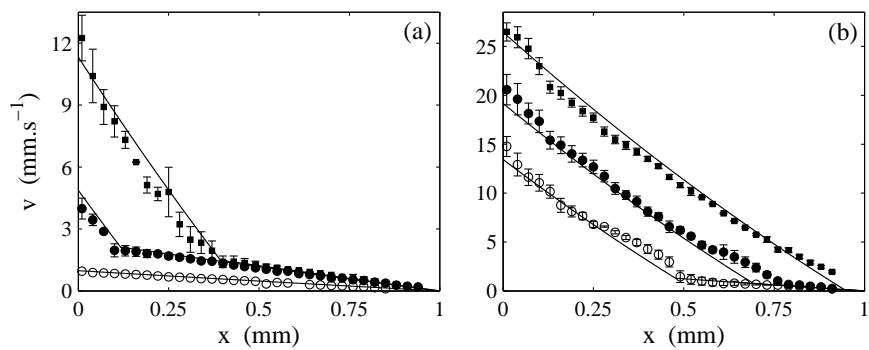}}
\caption{Time-averaged velocity profiles recorded using heterodyne DLS while measuring the flow curve of fig.~\ref{f.cpcl1} at (a) $\gammap=$ 1 ($\circ$), 5 ($\bullet$), and 12 ($\blacksquare$) s$^{-1}$ and (b) $\gammap=$ 15 ($\circ$), 22 ($\bullet$), and 28 ($\blacksquare$) s$^{-1}$. A highly sheared band grows from the inner rotating cylinder (at $x=0$) towards the stator (at $x=1$~mm). The solid lines correspond to the predictions of the simple gradient banding scenario with $\gammap_1=2.5$~s$^{-1}$, $\gammap_2=26$~s$^{-1}$, and $\sigma_c = 64.2$~Pa. Reprinted from ref.~\cite{Manneville:2004c}.}
\label{f.cpcl2}   
\end{figure}

\begin{figure}[htbp]
\scalebox{0.7}{\includegraphics{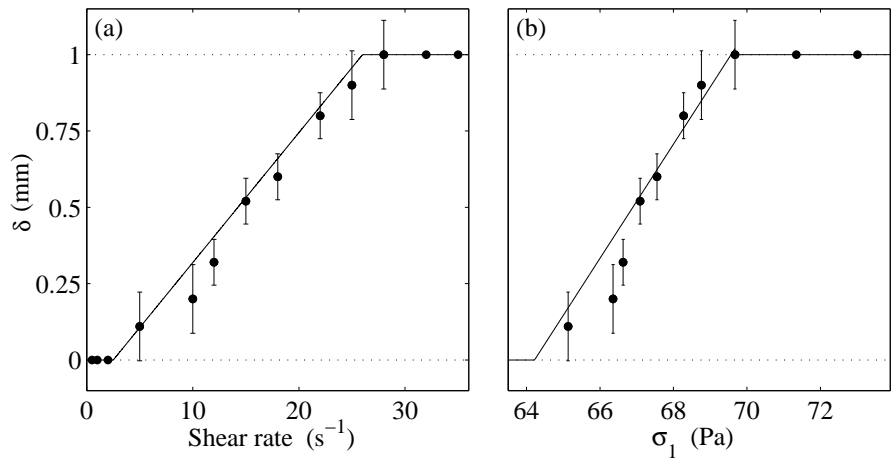}}
\caption{Width $\delta$ of the highly sheared band inferred from the velocity profiles of fig.~\ref{f.cpcl2}. (a) $\delta$ vs the engineering shear rate $\gammap$. (b) $\delta$ vs the local shear stress at the rotor $\sigma_1$. The solid lines correspond to the predictions of the simple gradient banding scenario with $\gammap_1=2.5$~s$^{-1}$, $\gammap_2=26$~s$^{-1}$, and $\sigma_c = 64.2$~Pa and show good agreement with the lever rule. Reprinted from ref.~\cite{Manneville:2004c}.}
\label{f.cpcl3}   
\end{figure}

Heterodyne DLS provided the first complete validations of the simple gradient banding scenario from time-averaged velocity profiles measured in the semidilute CPyCl--NaSal wormlike micellar system and in a lyotropic lamellar phase \cite{Manneville:2004c,Salmon:2003c,Salmon:2003d}. Figures~\ref{f.cpcl1}--\ref{f.cpcl3}  summarize the results obtained on the CPyCl--NaSal system at 6\%~wt. in 0.5~M brine \cite{Manneville:2004c,Salmon:2003c}. In the case of this transparent system, seeding the fluid with a small amount of nanometric latex particles was necessary to obtain enough light scattering. The same shear banding behaviour was found for time-averaged velocity profiles recorded under imposed shear rate during the layering transition from disordered to ordered onions, where seeding of the fluid was not necessary thanks to the weak scattering from the defects of the lamellar phase \cite{Manneville:2004c,Salmon:2003d}. The DLS technique was also applied to concentrated emulsions whose optical index contrast was finely tuned to get weakly scattering systems \cite{Salmon:2003a}. More recent versions of the heterodyne DLS setup have improved the temporal resolution to 0.01~s per velocity measurement and have been used to study shear banding in a supramolecular polymer solution \cite{vanderGucht:2006} and in Laponite suspensions \cite{Ianni:2007pp}.

\subsection{Ultrasonic velocimetry}
Ultrasound may be used as a nonintrusive probe in complex materials and does not require that the sample be optically transparent or quasi-transparent. In the Bordeaux group, ultrasonic velocimetry has been adapted to rheological experiments in Couette geometry and allows one to access velocity profiles every 0.02--2~s depending on the shear rate with a spatial resolution of about 40~$\mu$m \cite{Manneville:2004a}. This velocimetry technique, referred to as ultrasonic speckle velocimetry (USV) or ultrasound velocity profiling (UVP), is based on the interaction between a short high-frequency ultrasonic pulse and particles suspended in the fluid. In some cases, the fluid microstructure may scatter ultrasound efficiently enough so that seeding the sample with particles is not needed \cite{Manneville:2005}.

\begin{figure}[htbp]
\scalebox{0.8}{\includegraphics{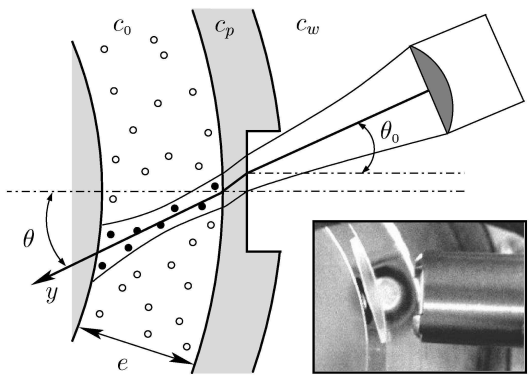}}
\caption{USV setup. $c_0$, $c_p$, and $c_w$ stand
for the speed of sound in the fluid under study, in Plexiglas, and in water
respectively. The inset shows a picture of the transducer and
the stator as seen from above once the rotor has been removed. Reprinted from ref.~\cite{Manneville:2004a}.}
\label{f.usv}   
\end{figure}

\begin{figure}[htbp]
\scalebox{1}{\includegraphics{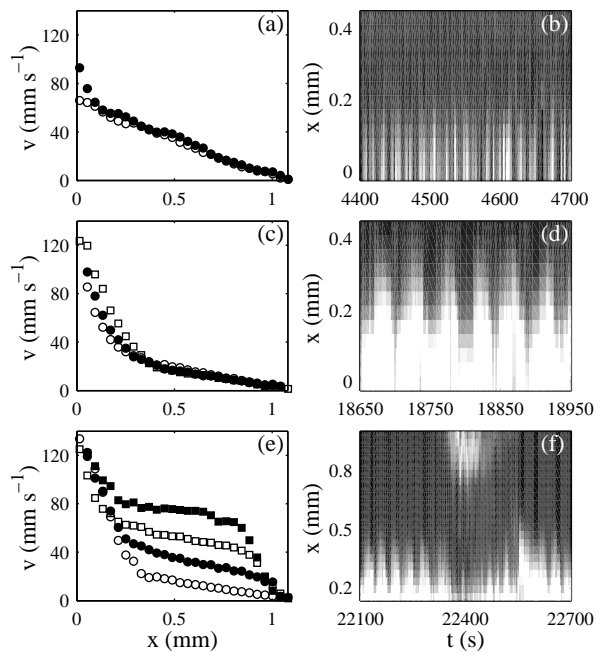}}
\caption{Spatiotemporal behaviours observed in a 20\%~wt. CTAB solution in D$_2$O under an imposed engineering shear rate $\gammap=186$~s$^{-1}$.
(a)-(b) Intermittent apparition of a highly sheared band at the rotor.
(c)-(d) Oscillations of the position of the interface between shear bands.
(e)-(f) Nucleation of a second highly sheared band at the stator.
(a), (c), and (e) present some individual velocity profiles while (b), (d), and (f) show  spatiotemporal diagrams of the local shear rate $\gammap(x,t)$ inferred from the local velocity measurements. Reprinted from ref.~\cite{Becu:2004}.}
\label{f.ctab1}   
\end{figure}

\begin{figure}[htbp]
\scalebox{1}{\includegraphics{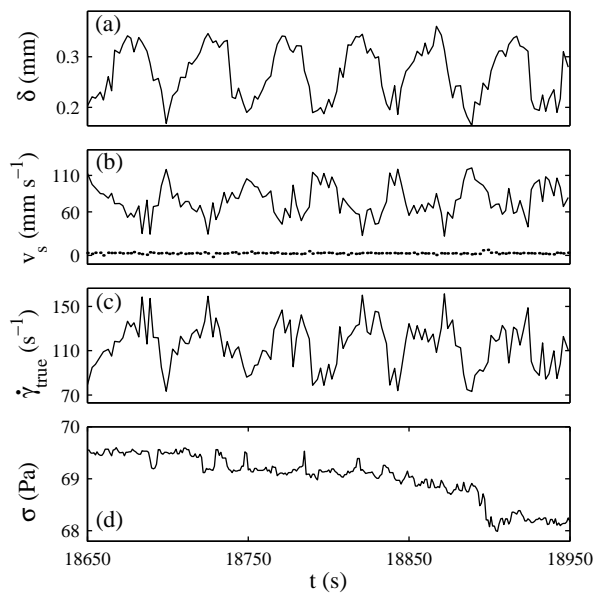}}
\caption{Temporal fluctuations of (a) the position $\delta(t)$ of the interface between the two shear bands, (b) the slip velocities at the rotor $v_{s1}(t)$ (solid line) and at the stator $v_{s2}(t)$ (dotted line), (c) the true shear rate $\gammap_{\rm true}(t)$, and (d) the temporal response of the engineering shear stress $\sigma(t)$ recorded during the oscillations of Fig.~\ref{f.ctab1}(c)-(d). The imposed engineering shear rate is $\gammap=186$~s$^{-1}$. Reprinted from ref.~\cite{Becu:2004}.}
\label{f.ctab2}   
\end{figure}

Figure~\ref{f.usv} shows the relative arrangement of the Couette cell and the ultrasonic transducer that works both in emission and reception at a central frequency of 36~MHz. Backscattered pressure signals that result from successive pulses sent by the transducer are recorded as a function of time. These signals constitute an ultrasonic ``speckle'' that is directly linked to the spatial distribution of the scatterers along the acoustic beam (as long as one remains in the single scattering regime). By cross-correlating two successive speckle signals over small time windows, one can measure the displacements of the scatterers at various positions along the beam. A calibration procedure in a Newtonian fluid is used to measure the incidence angle $\theta$ shown in fig.~\ref{f.usv} from which quantitative estimates of the tangential velocity are inferred. Good statistical convergence is obtained by averaging over typically 1000 pulses (see ref.~\cite{Manneville:2004a} for more details). 

Thanks to its fast temporal resolution, USV has been used to get better insight into the spatiotemporal fluctuations of shear-banded flows that were first detected by DLS or NMR measurements. In particular USV experiments during the layering transition from disordered to ordered onions in multilamellar vesicles have evidenced a complex interplay between wall slip and bulk dynamics that leads to the failure of the ``instantaneous'' lever rule (see also figs.~\ref{f.onion1} and \ref{f.onion2} below) \cite{Manneville:2004b}. Similarly it was shown that shear banding in a concentrated wormlike micellar solution involves strong wall slip, oscillations of the interface position between shear bands, and transient nucleation of three-banded flows \cite{Becu:2007,Becu:2004}. Figures~\ref{f.ctab1} and \ref{f.ctab2} illustrate the typical spatiotemporal behaviours observed in a CTAB solution at 20\%~wt. in D$_2$O. Recent applications of the USV technique include shear-thickening wormlike micelles \cite{Decruppe:2006,Herle:2007_pp}, concentrated emulsions \cite{Becu:2006}, triblock copolymers \cite{Manneville:2007}, and organogels \cite{Grondin:2007_pp}.

\subsection{Comparison between velocimetry techniques}
In order to confront the various velocimetry techniques described in this section, table~\ref{t.velo} gathers the characteristics and typical resolutions obtained through NMR, PTV, DLS, and USV respectively. It also lists the main advantages and drawbacks of each technique, so that the experimentalist interested in velocity measurements may find the most suitable way to assess a complex system under shear depending on its physical properties and on the desired resolutions.

\begin{table}[ht]
\begin{center}
\begin{tabular}{||p{2cm}||p{2.9cm}|p{2.9cm}|p{2.9cm}|p{2.9cm}||}
\hline\hline
 & {\centering{\smallskip NMR\\}} & {\centering{\smallskip PTV\\}} & {\centering{\smallskip DLS\\}} & {\centering{\smallskip USV\\}} \\
\hline\hline
{\centering{\smallskip Spatial resolution\\}} & {\centering{\smallskip 50--100~$\mu$m\\}} & {\centering{\smallskip 1--10~$\mu$m\\}} & {\centering{\smallskip 50--100~$\mu$m\\}} & {\centering{\smallskip 40~$\mu$m\\}} \\
\hline
{\centering{\smallskip Temporal resolution\\}} & {\centering{\smallskip 1~s per profile\\}} & {\centering{\smallskip 2.5~ms--0.1~s per profile\\}} & {\centering{\smallskip 10~ms--1~s per point\\10--100~s per profile\\}} & {\centering{\smallskip 20~ms--2~s per profile\\}}\\
\hline
{\centering{\smallskip System requirements\\}} & {\centering{\smallskip none\\}} & {\centering{\smallskip optically transparent\\+ tracer particles\\}} & {\centering{\smallskip weak light scattering\\}} & {\centering{\smallskip weak ultrasound scattering\\}} \\
\hline
{\centering{\smallskip Advantages\\}} &{\centering{\smallskip- works with turbid media\\- may also yield structural~information\\- may yield 2D images\\}}& {\centering{\smallskip- excellent spatial and temporal resolutions\\- yields 2D velocity vector\\}} & {\centering{\smallskip- may also yield structural information through SALS}} & {\centering{\smallskip- works with turbid media\\- easy to implement on a rheometer\\}}\\
\hline
{\centering{\smallskip Drawbacks\\}} & {\centering{\smallskip- hard to achieve both high spatial and temporal resolutions\\- no simultaneous measurement of rheological data\\- costly technique}} & {\centering{\smallskip- requires transparent system\\- requires seeding\\}} & {\centering{\smallskip- requires seeding if optically transparent\\- pointlike measurements\\}} & {\centering{\smallskip- requires seeding if acoustically transparent\\}}\\
\hline\hline
\end{tabular}
\end{center}
\caption{\label{t.velo}Comparison between the current velocimetry techniques used for investigating shear-banded flows.}
\end{table}

\pagebreak
\section{Open questions and future directions for research}
\label{questions}

Examples discussed above have shown (i) the validity of the simple gradient banding scenario in a few systems when the stationary state is reached and when time-averaged velocity profiles are considered (see figs.~\ref{f.cpcl1}--\ref{f.cpcl3}) and (ii) complex spatiotemporal phenomena when time-resolved measurements are performed (see figs.~\ref{f.ctab1} and \ref{f.ctab2}). More generally, from recent measurements using the various techniques described previously, a number of complications seem to occur that infirm the simple shear banding scenario even on time-average. In this section we wish to further discuss a few open questions in the domain of shear banding. Without entering too much into the details of specific systems, we shall try to raise important issues that require further experimental investigations.

\subsection{The nature and the dynamics of the shear banding transition}

In connection with theoretical approaches of shear banding, various experimental studies have been conducted to investigate the mechanism of the shear banding transition. More precisely, in wormlike micellar solutions, it appears that some semidilute solutions nicely follow a mechanical instability scenario with a ``top-jumping'' scenario as in the original picture of Cates {\it et al.} \cite{Spenley:1993}, whereas more concentrated solutions rather present a shear-induced phase transition characterized by metastable states and sigmoidal transients \cite{Berret:1997b,Berret:1997a,Berret:1994b,Grand:1997}. The comparison between semidilute and concentrated systems has been addressed in the review article \cite{Berret:2005}. Refs.~\cite{Hu:2005,Miller:2007} also provide careful investigations of the kinetics and mechanism of shear banding in the semidilute CPyCl--NaSal system using PTV. Recent modelling of shear-banded flows seems to have somehow settled this debate of ``mechanical instability vs phase transition'' since a smooth crossover between the two scenarii is expected when one accounts for flow--concentration coupling. The reader is referred to theoretical companion reviews in the same volume for more details.

A major advance in modelling the experimental observations on shear banding has been to include nonlocal diffusive terms in the evolution equations for the rheological variables \cite{Dhont:1999,Olmsted:1997,Radulescu:1999,Yuan:1999}. Using the so-called ``diffusive Johnson-Segalman model,'' it was shown that such diffusive terms account for metastable states, history independence, unicity of the selected stress $\sigma_c$, and the presence of only two shear bands with a single interface \cite{Olmsted:2000,Radulescu:2000}. A current experimental challenge lies in measuring the diffusion coefficient $\cal{D}$ and the corresponding correlation length $\zeta$. Using rheo-optics, Radulescu {\it et al.} \cite{Radulescu:2003} found $\zeta\simeq 40$--120~nm in CTAB--NaNO$_3$ and CTAB--KBr samples, close to the mesh size of these micellar networks, whereas later measurements on the CPyCl--NaSal system using superposition rheology and ultrasonic velocimetry \cite{Ballesta:2007} and particle tracking in microchannels \cite{Masselon:2007_pp} rather point to much larger values $\zeta\simeq 2$--20~$\mu$m. Such a discrepancy in similar semidilute systems is striking and remains to be elucidated, e.g., through more experiments on a wider variety of systems.

\subsection{Spatiotemporal fluctuations and interface instability}

In some cases, strong temporal fluctuations of both the rheological variables and the flow field have been measured, which can even make it hard to define a steady-state. Such fluctuations were first evidenced indirectly through spatially-resolved velocity distributions using rheo-NMR \cite{Holmes:2003a} then directly with time-resolved NMR \cite{Lopez:2004} and ultrasonic velocimetry \cite{Becu:2004} (see figs.~\ref{f.ctab1} and \ref{f.ctab2}). These observations have triggered new theoretical works on the stability of shear-banded flows and rheochaos in both the shear-thinning and shear-thickening cases \cite{Aradian:2005,Aradian:2006,Fielding:2005,Fielding:2004,Fielding:2006}. 

Moreover recent flow visualizations have evidenced an instability of the interface between shear bands along the vorticity direction in the CTAB--NaNO$_3$ wormlike micellar system and different spatiotemporal regimes were found \cite{Lerouge:2006}. The same kind of instability may explain the interface oscillations detected in other systems such as those of fig.~\ref{f.ctab1}(c)-(d). More importantly, these experimental results question the validity of the assumption of purely tangential flows. Indeed hints for three-dimensional flows have been reported from ultrasonic measurements \cite{Becu:2007,Manneville:2007}. It remains to be checked whether such unstable flows result from an instability of the interface or from other instabilities such as elastic instabilities. Direct evidence for secondary flows and/or radial velocity fluctuations have also been obtained using PTV in granular materials \cite{Mueth:2003,Mueth:2000}, two-dimensional foams \cite{Debregeas:2001,Kabla:2003}, and rodlike viruses \cite{Kang:2006}. Future technical effort should probably focus on whether the NMR or USV techniques may be extended to provide three-dimensional data with a fast enough temporal resolution.

\begin{figure}[htbp]
\scalebox{1}{\includegraphics{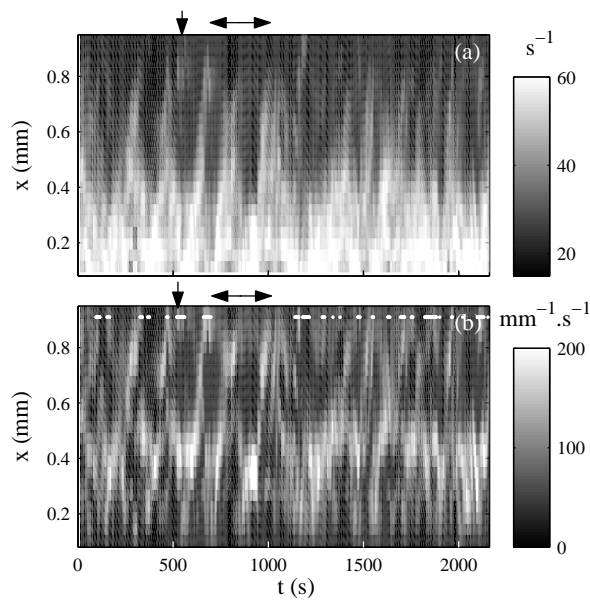}}
\caption{Shear banding in the SDS--octanol in brine system under an imposed engineering shear stress $\sigma=19$~Pa. This lamellar phase presents a shear-induced layering transition from disordered to ordered onions. (a) Local shear rate $\gammap(x,t)$. (b) Spatial derivative of the local shear rate: $\partial\gammap(x,t)/\partial x$. This derivative gives an indication for the position of the interface between shear bands. White lines and dots indicate time intervals when three (or more) shear bands are detected. Reprinted from ref.~\cite{Manneville:2004b}.}
\label{f.onion1}   
\end{figure}

Another interesting issue for future research would be to explain the origin of the time scales involved in rheochaotic fluctuations and their order of magnitude (fractions of a second to minutes). A related question is whether such temporal fluctuations and instabilities are inherent to shear-banded flows and whether they may have some ``universal'' features regarding shear-induced transitions. Indeed chaotic-like fluctuations were observed in various materials of widely different microstructures \cite{Bandyopadhyay:2000,Bandyopadhyay:2001,Ganapathy:2006,Pimenta:2006,Pujolle:2003,Salmon:2002}. Figure~\ref{f.onion1} shows an example of the complex flow encountered in a chaotic-like regime during the layering transition from disorder to ordered onions in the SDS--octanol in brine lamellar phase. As in fig.~\ref{f.ctab1}(e)-(f), transient three-banded states are detected, which clearly violates any simple ``instantaneous'' shear banding scenario and raises even more theoretical questions.

\subsection{The importance of wall slip and boundary conditions in shear-banded flows}

Another challenge for both experimental and theoretical studies is to investigate the effect of wall slip and its coupling to a bulk shear-banded flow. Indeed apparent slip at the boundaries is ubiquitous in the rheology of complex fluids \cite{Barnes:1995}. Although roughening the walls may help to prevent the fluid from slipping at the walls, it is sometimes difficult to check whether slip is actually present or not from rheological data alone. However extrapolating local velocity measurements at the cell boundaries easily provides quantitative estimates of slip velocities. Knowing the slip velocities, one can infer the ``true'' shear rate (see fig.~\ref{f.ctab2}(b) and (c)) and construct the ``true'' flow curve which may significantly differ from the engineering flow curve.

\begin{figure}[htbp]
\scalebox{1}{\includegraphics{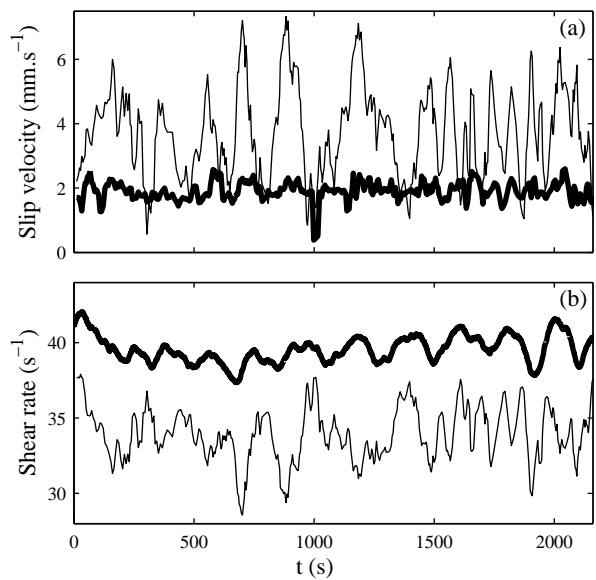}}
\caption{(a) Slip velocities at the rotor $v_{s1}(t)$ (thin line) and at the stator $v_{s2}(t)$ (thick line) inferred from the measurements of fig.~\ref{f.onion1}.
(b) Engineering shear rate $\gammap(t)$ (thick line) and true shear rate $\gammap_{\rm true}(t)$ (thin line). Reprinted from ref.~\cite{Manneville:2004b}.}
\label{f.onion2}   
\end{figure}

The coupling between wall slip and, {\it e.g.}, interface dynamics in a shear-banded flow remains an open issue. As shown in fig.~\ref{f.onion2}(a) in the case of a lamellar phase undergoing the transition from disordered to ordered onions, wall slip at the rotor and at the stator may be very different. In that case the ordered onions (close to the rotor) present much higher slip velocities than the disordered onions (close to the stator). Moreover, once wall slip is taken into account, one finds that the ``true'' shear rate experienced by the sample is 20\% smaller than the engineering shear rate and presents very large fluctuations that are strongly correlated to the position of the interface between shear bands \cite{Manneville:2004b}. Another example of a strong dissymmetry in slip velocities at the rotor and at the stator is found in fig.~\ref{f.ctab2}(b) in the case of concentrated wormlike micelles. Such results should prompt experimentalists to investigate in more details the structure and the rheology of the slip layers. 

\begin{figure}[htbp]
\begin{center}
\scalebox{1}{\includegraphics{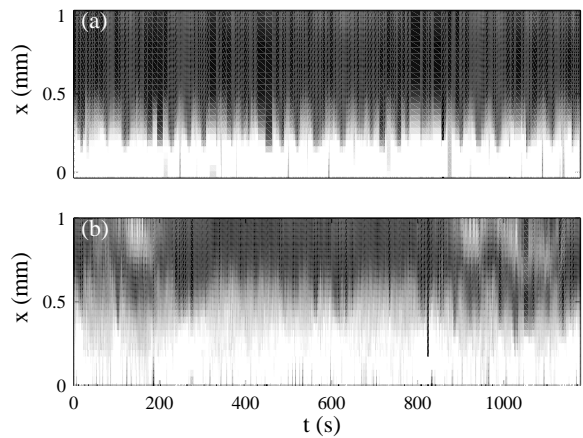}}
\end{center}
\caption{Local shear rate $\gammap(x,t)$ recorded in a 20\%~wt. CTAB solution in D$_2$O under an imposed engineering shear rate $\gammap=400$~s$^{-1}$ (a) in a smooth Couette cell and (b) in a rough Couette cell. Reprinted from ref.~\cite{Becu:2007}.}
\label{f.ctab3}
\end{figure}

Figure~\ref{f.ctab3} provides another example of the importance of wall slip in the shear-banded flow of concentrated CTAB--D$_2$O wormlike micelles. Time-resolved velocity profiles were measured in both a smooth Plexiglas Couette cell and in a roughened (sand-blasted) cell with the same dimensions and for the same imposed engineering shear rate. Although the ``true'' shear rates are not the same due to the reduction of wall slip in the rough cell (as can be seen from the average position of the interface between shear bands), fig.~\ref{f.ctab3} shows that the bulk dynamics are deeply modified when the boundary conditions are changed. The large periodic oscillations of the position of the interface between shear bands recorded in the smooth geometry (fig.~\ref{f.ctab3}(a)) are replaced by smaller and more erratic fluctuations in the rough Couette cell, where a second highly sheared band is also detected transiently in the vicinity of the stator (fig.~\ref{f.ctab3}(b)).

These observations on wall slip raise the more general question of whether shear banding may be affected by boundary conditions. This issue has been addressed recently in theoretical and numerical studies \cite{Adams:2007,Heidenreich:2007,Rossi:2006} and appears as a major challenge for future research. In particular, the interplay between the boundary conditions and the geometry could provide some clues to explain the stationary three-banded flow observed in a triblock copolymer solution \cite{Manneville:2007}. Studying the effect of confinement on shear-banded flows, {\it e.g.}, through the use of microfluidics, should also yield new insights on wall slip and finite-size effects \cite{Degre:2006,Goyon:2007_pp,Masselon:2007_pp}. Finally, from a purely technical point of view, one may further speculate that interesting answers on the physics of wall slip and its coupling to shear banding could come from (i) developing techniques with submicron resolution to, {\it e.g.}, directly estimate the thickness and the structure of the slip layers and (ii) better characterization and control of the wall properties using, {\it e.g.}, nanopatterned surfaces in rheological experiments \cite{Cottin:2003}.

\subsection{The ``particular'' case of soft glassy materials}

As recalled in the introduction, the solid--liquid transition (or ``unjamming'' transition) in soft glassy materials under shear may be seen as a particular case of shear banding. However, since it involves vanishingly small shear rates as well as generally thixotropic materials, such a shear-induced transition remains the subject of intense debate. In particular the question of whether this transition is continuous (and can be modelled using a standard yield stress fluid such as a Bingham or a Herschel-Bulkley fluid) or discontinuous (and involves shear localization above some critical shear rate) is still open. Another issue concerns the interactions between particles in concentrated systems and their effect on shear banding phenomena. The various velocimetry techniques listed in section~\ref{velo} have been used on soft glassy materials. Recent works are mentionned in the corresponding sections above. In our opinion understanding shear banding in soft glassy systems and elucidating unjamming mechanisms constitute some of the most challenging research directions in the field.

\section{Conclusion}

In summary this paper has provided a review of state-of-the-art experimental techniques to assess shear banding. Both local strucural and velocity measurements have recently shed new light on shear-banded flows, confirming the predicted simple behaviour in some cases but also evidencing unexpected features in other cases, mainly through transient measurements with high temporal resolution. The latest experimental studies in the field now prompt us to further technical developments focusing on the possibility of three-dimensional instabilities specific of shear-banded flows, on more precise investigations of wall slip and its interplay with shear banding, and on yielding in soft glassy materials.

\vspace{0.5cm}
The author wish to thank A.~Aradian, P.~Ballesta, L.~B\'ecu, A.~Colin, J.-P.~Decruppe, S.~Fielding, P.~Fischer, O.~Greffier, V.~Herle, S.~Lerouge, P.~Lettinga, F.~Molino, P.~Olmsted, G.~Ovarlez, J.-B.~Salmon, and F.~Schosseler for collaboration and fruitful discussions.

\end{document}